\documentclass[a4paper]{spie}  
\usepackage[]{graphicx}
\usepackage{textcomp}
\usepackage{amsmath}
\usepackage{booktabs}

\title{Results of the simulations of the petal/lens as part of the LAUE project} 

\author{V.~Valsan\supit{a,b}, F.~Frontera\supit{a}, E.~Virgilli\supit{a}, V.~Liccardo\supit{a,b}, 
E.~Caroli\supit{c}, J.B.~Stephen\supit{c}
\skiplinehalf
\supit{a} \small\textit{Physics and Earth Science Department, University of Ferrara, Via Saragat 1, 44122 Ferrara, Italy};\\
\supit{b} \small\textit{Universit\'e de Nice Sophia-Antipolis, Parc Valrose, 06108 Nice Cedex 2, France};\\
\supit{c} \small\textit{IASF-INAF via P.Gobetti, Bologna - Italy}.
}

\authorinfo{Send correspondence to:\\
valsan@fe.infn.it, frontera@fe.infn.it, virgilli@fe.infn.it, liccardo@fe.infn.it
}

\begin{document} 

\maketitle

\begin{abstract}

In the context of the LAUE project for focusing hard X-/gamma rays, a petal of the complete lens is being 
assembled at the LARIX facility in the Department of Physics and Earth Science of the University of Ferrara. 
The lens petal structure is composed of bent Germanium and Gallium Arsenide crystals in 
transmission geometry. We present the expectations derived from a mathematical model of the lens petal. 
The extension of the model for the complete LAUE project in the 90 -- 600 keV 
energy range will be discussed as well.
A quantitative analysis of the results of these simulations is also presented.

\end{abstract}


\section{INTRODUCTION}
\label{sec:intro}

We propose a Laue lens\cite{Frontera12} as a new focusing instrument 
in the soft gamma--ray band (80--600 keV) in order to 
answer most of the unresolved scientific open issues in soft gamma-ray astronomy \cite{Frontera13}. 
(For an exhaustive review of Laue lenses see Ref.~[\citenum{Frontera11}]). 
Lens prototypes have already been successfully built in the 
LARIX facility \cite{Loffredo05} of the University of Ferrara\cite{Virgilli11,Frontera08,Ferrari09,Frontera07}.
Preliminary simulation results  of a petal made with Ge(111) and GaAs(111) was reported in [\citenum{Valsan12}].
The petal that is being built is made up of both GaAs(220)
and Ge(111) crystal tiles\cite{Frontera13, Liccardo13,Virgilli13}.
The selection of the crystals and their characterization is presented in [\citenum{Liccardo12}]. 
Here we report the final simulation and performance results of this lens petal and of a lens made of petals 
like that is being assembled of either Ge(111) or GaAs(220).

\section{Modelling the petal structure}

\begin{figure}[h!]
  \centering
  \includegraphics[scale=0.5,keepaspectratio=true]{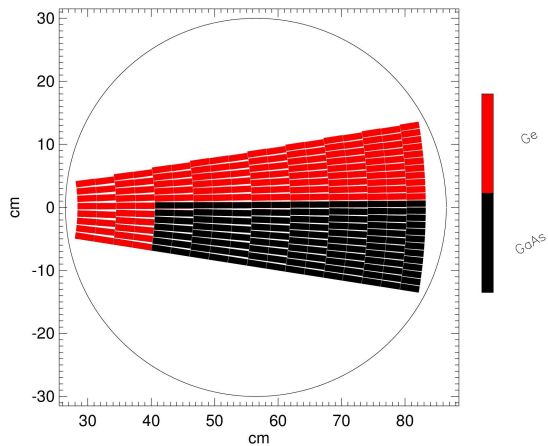} 
  \caption{An illustration of the petal structure (that is being build in the LARIX facility)
  and the position of each crystal tile over it.}
  \label{fig:petal_structure}
\end{figure}

The petal structure is modeled in such a way as to diffract photons
in the energy band of 90 keV -- 304 keV. This pass band is the actual energy band of the photons currently 
available in the LARIX facility.  The crystals will be positioned and glued on the structure so 
as to focus the diffracted beam at a distance of 20 meters. 
Figure~\ref{fig:petal_structure} illustrates the petal structure and also the respective positions of
each crystal tile. Table~\ref{tab:Petal_LARIX} shows main parameters of this petal.

\begin{table}[h]
  \begin{center}  
    \begin{tabular}{ l l l l }
    \toprule
    Parameter 		& \multicolumn{3}{c}{Value} 	\\ \cmidrule(r){2-4}
			& section of GaAs(220) 		& section of Ge(111)	& Entire petal	\\ \midrule
    Focal length	& 20 meters 			& 20 meters 		& 20 meters	\\
    Energy range	& 148 -- 304 keV		& 90 -- 267 keV		& 90 -- 304 keV	\\
    No. of Rings	& 14				& 18			& 18		\\
    Innermost radius	& 40.66 cm			& 28.40 cm		& 28.40 cm	\\
    Outermost radius	& 83.47 cm 			& 83.47 cm 		& 83.47 cm	\\
    No. of crystal tiles & 119				& 155	 		& 274		\\
    Crystal size 	& 30 $\times$ 10  $\times$ 2 mm$^3$& 30 $\times$ 10 $\times$ 2mm$^3$	& 30 $\times$ 10 $\times$ 2 mm$^3$	\\
    Crystal mass (total) & 2.5 g $\times$ 119 = 297.5 g		& 2.07 g $\times$ 155 = 320.85 g & 618.35 g		\\
    \bottomrule
    \end{tabular}
    \newline
    \caption{Parameters of the petal that is being build in the LARIX facility.}
    \label{tab:Petal_LARIX}
  \end{center}	
\end{table}

\section{Simulating the petal PSF}

The Point Spread Function (PSF) gives the spatial distribution of the reflected photons in the focal plane 
when a point source at infinity is incident over the lens or petal.

The resulting PSF has been simulated taking into account all types of misalignment and distortion effects 
of every single crystal. 
A slight misalignment of a crystal from its nominal position creates a distortion 
in the PSF, therefore a precise positioning of each single crystal on the lens is crucial.  

Another factor which affects the  PSF is the  distortion in the curvature of the crystals. 
The petal will be made of GaAs(220) and Ge(111) crystals curved with a radius of 40 meter. 
Any distortion in the curvature radius will also affect the PSF.
The lens petal which has been simulated, is separately made up of Ge(111) and GaAs(220) crystal tiles, 
with an energy passband from $\sim$90 keV to $\sim$300 keV. 

\begin{figure}[!ht]
\centering
\begin{minipage}{.5\textwidth}
    \centering
    \includegraphics[scale=0.4]{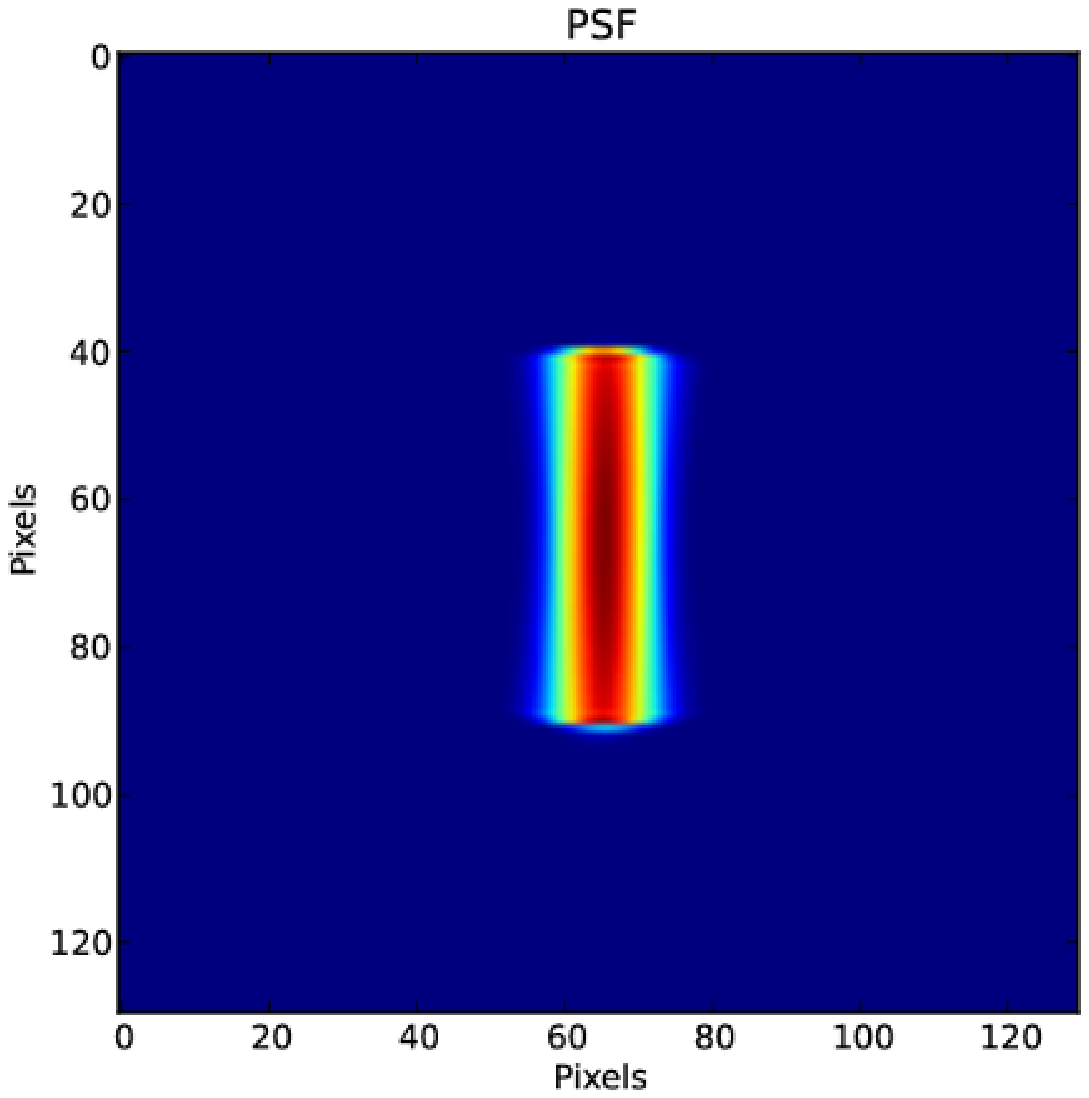}
    
\end{minipage}%
\begin{minipage}{.5\textwidth}
    \centering
    \includegraphics[scale=0.6]{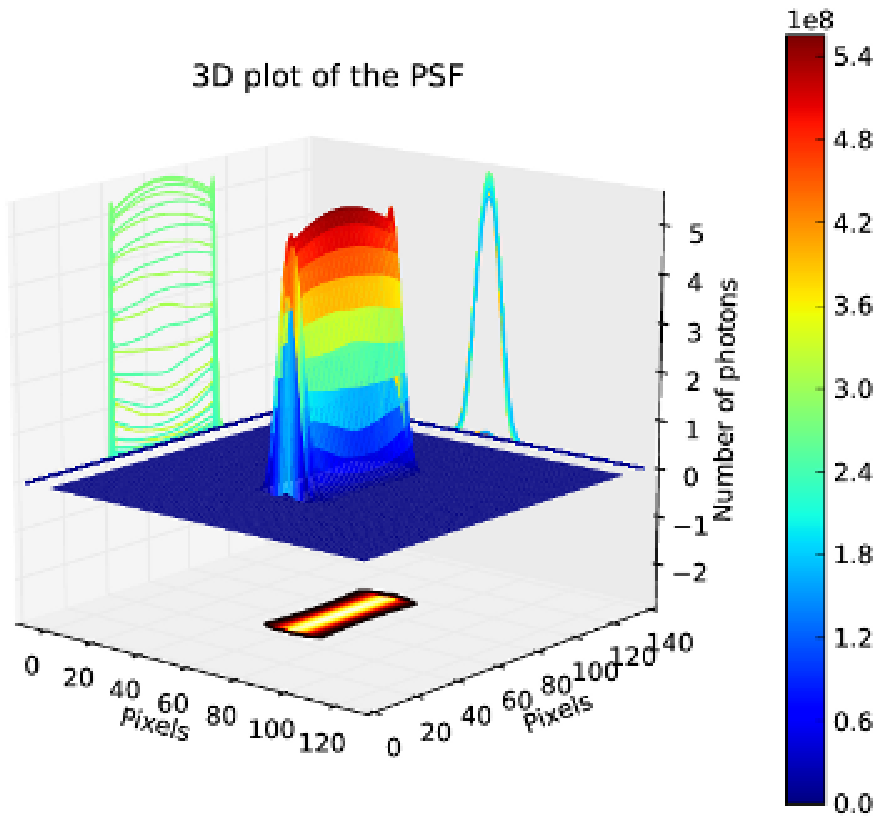} 
    
\end{minipage}
\caption{PSF of the petal made with GaAs(220) without any misalignment errors in the positioning of the 
crystal tiles having no radial distortion.}
\label{fig:P_GaAs_M00R0}
\end{figure}

\begin{figure}[!ht]
\centering
\begin{minipage}{.5\textwidth}
    \centering
    \includegraphics[scale=0.4]{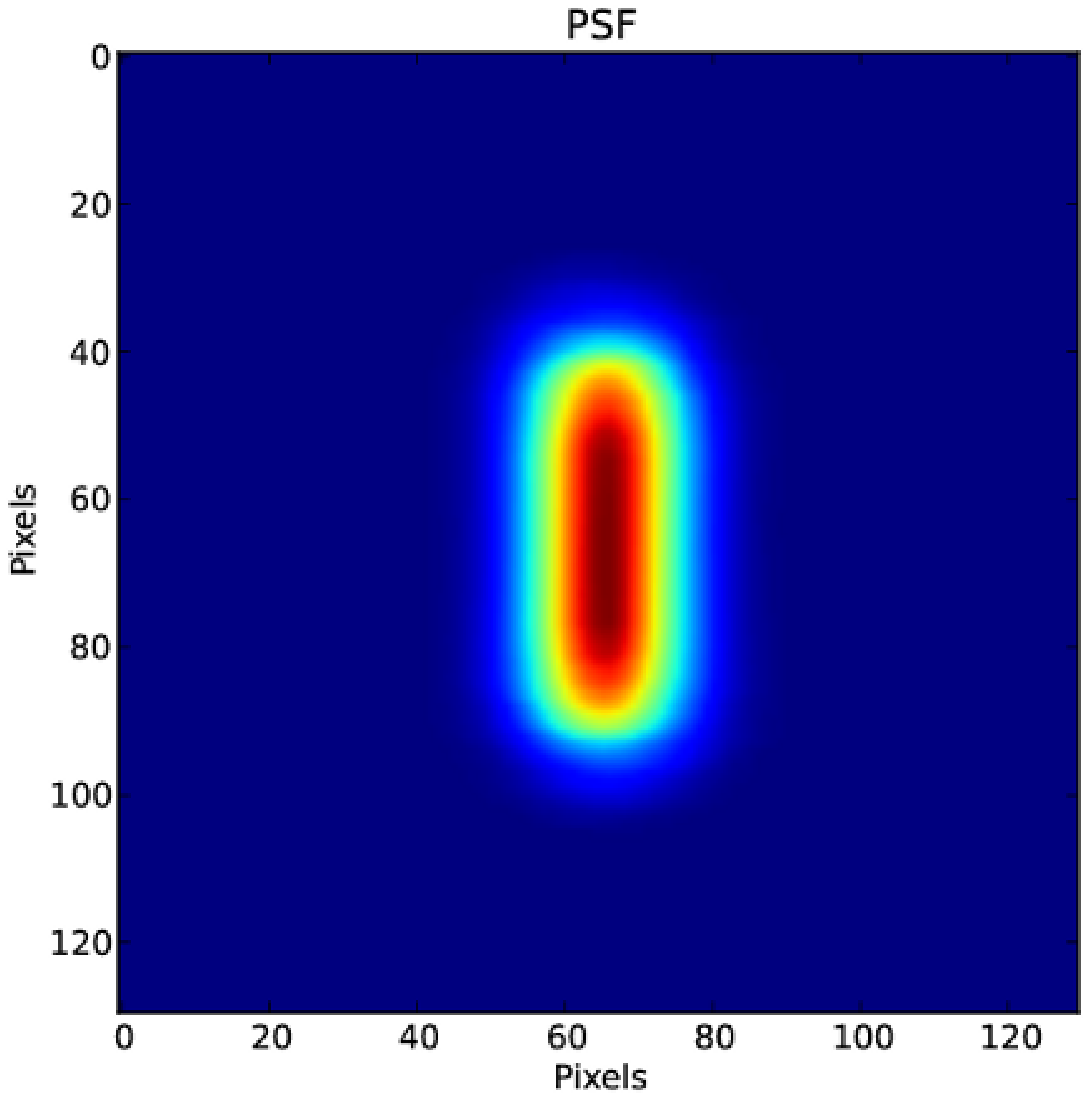}
    
\end{minipage}%
\begin{minipage}{.5\textwidth}
    \centering
    \includegraphics[scale=0.6]{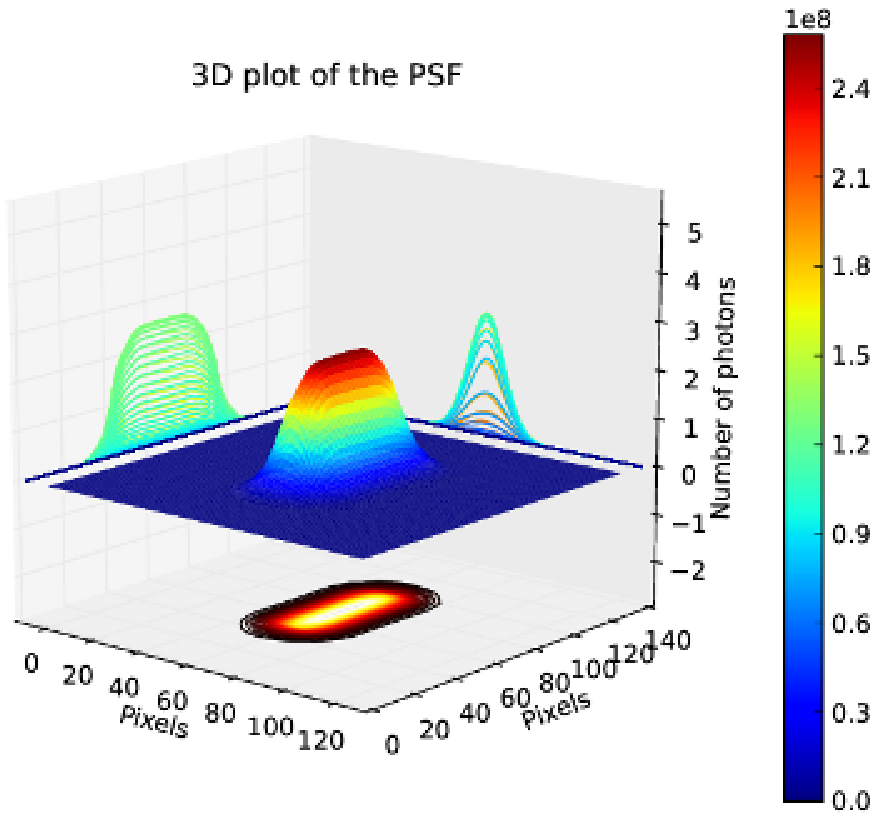} 
    
\end{minipage}
\caption{PSF of the petal made with GaAs(220) with a maximum misalignment of 30 arcsec in the 
positioning of the crystal tiles having a maximum radial distortion of 6 meters.}
\label{fig:P_GaAs_M30R6}
\end{figure}

\begin{figure}[!ht]
\centering
\begin{minipage}{.5\textwidth}
    \centering
    \includegraphics[scale=0.4]{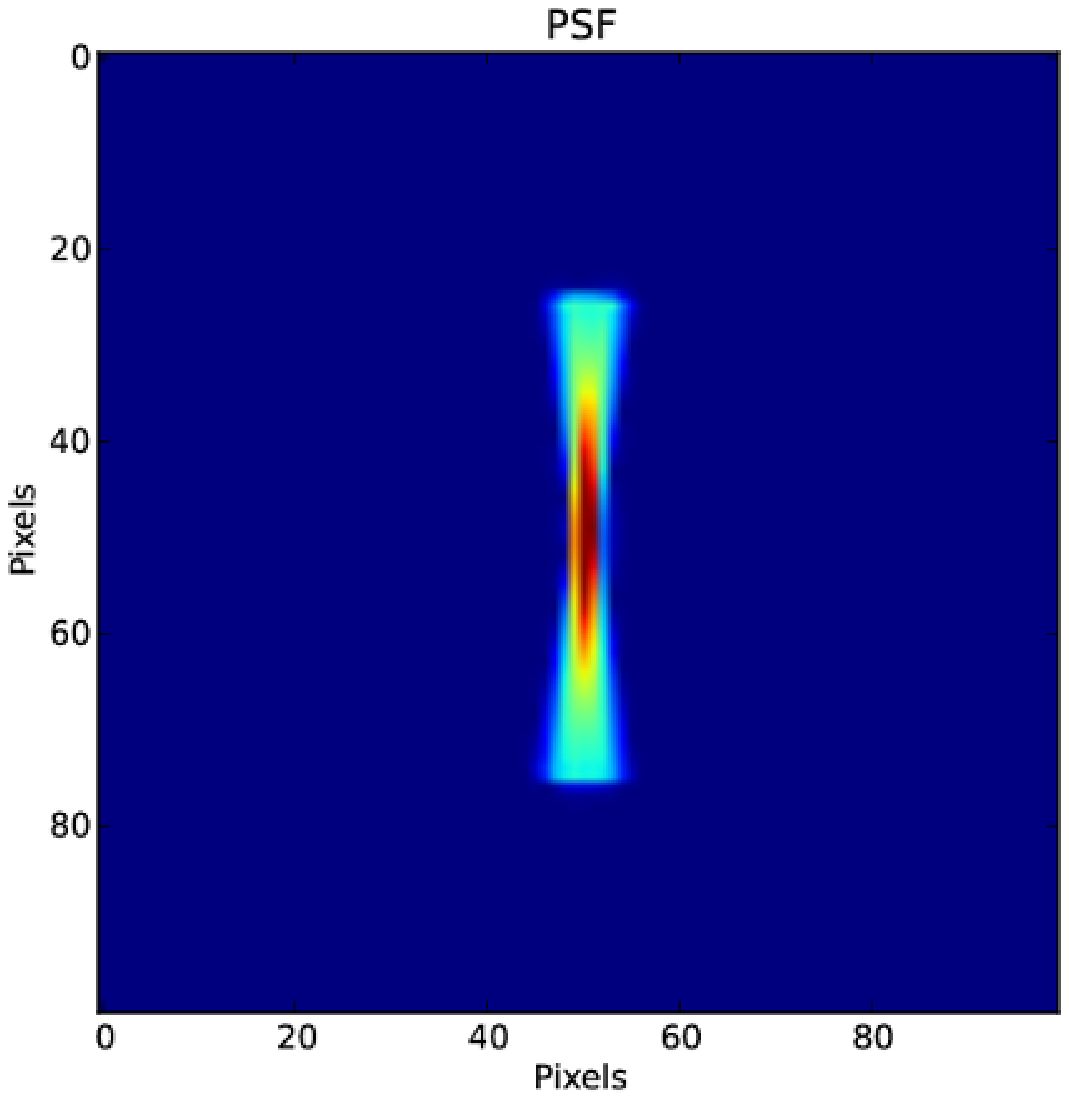}
    
\end{minipage}%
\begin{minipage}{.5\textwidth}
    \centering
    \includegraphics[scale=0.6]{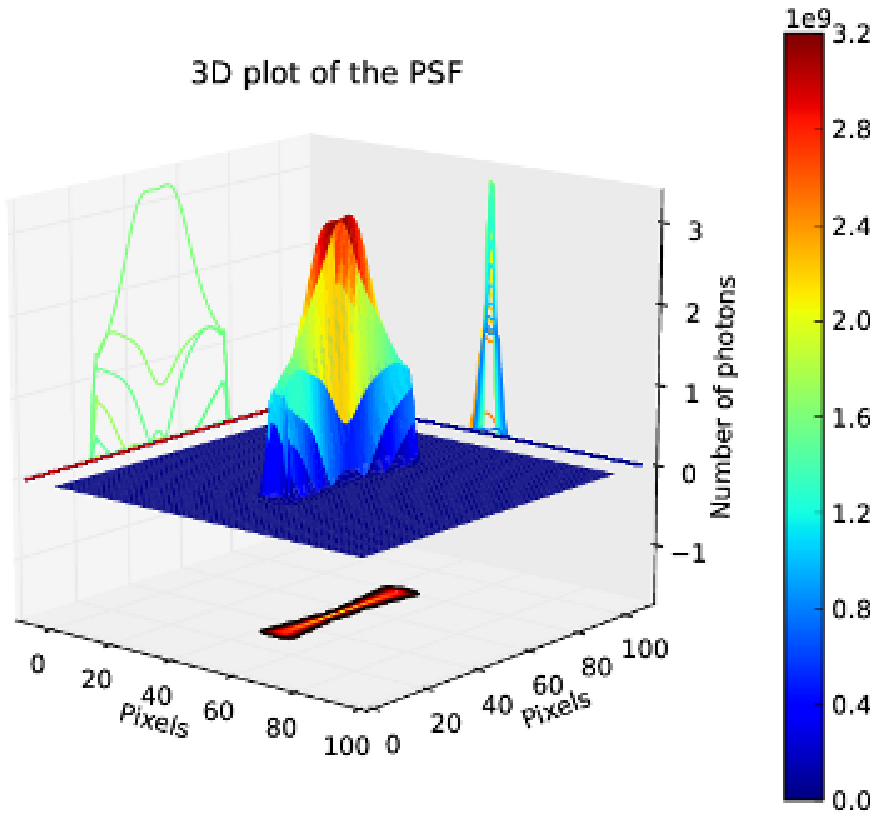} 
    
\end{minipage}
\caption{PSF of the petal made with Ge(111) without any misalignment errors in the positioning of the 
crystal tiles having no radial distortion.}
\label{fig:P_Ge_M00R0}
\end{figure}

\begin{figure}[!ht]
\centering
\begin{minipage}{.5\textwidth}
    \centering
    \includegraphics[scale=0.4]{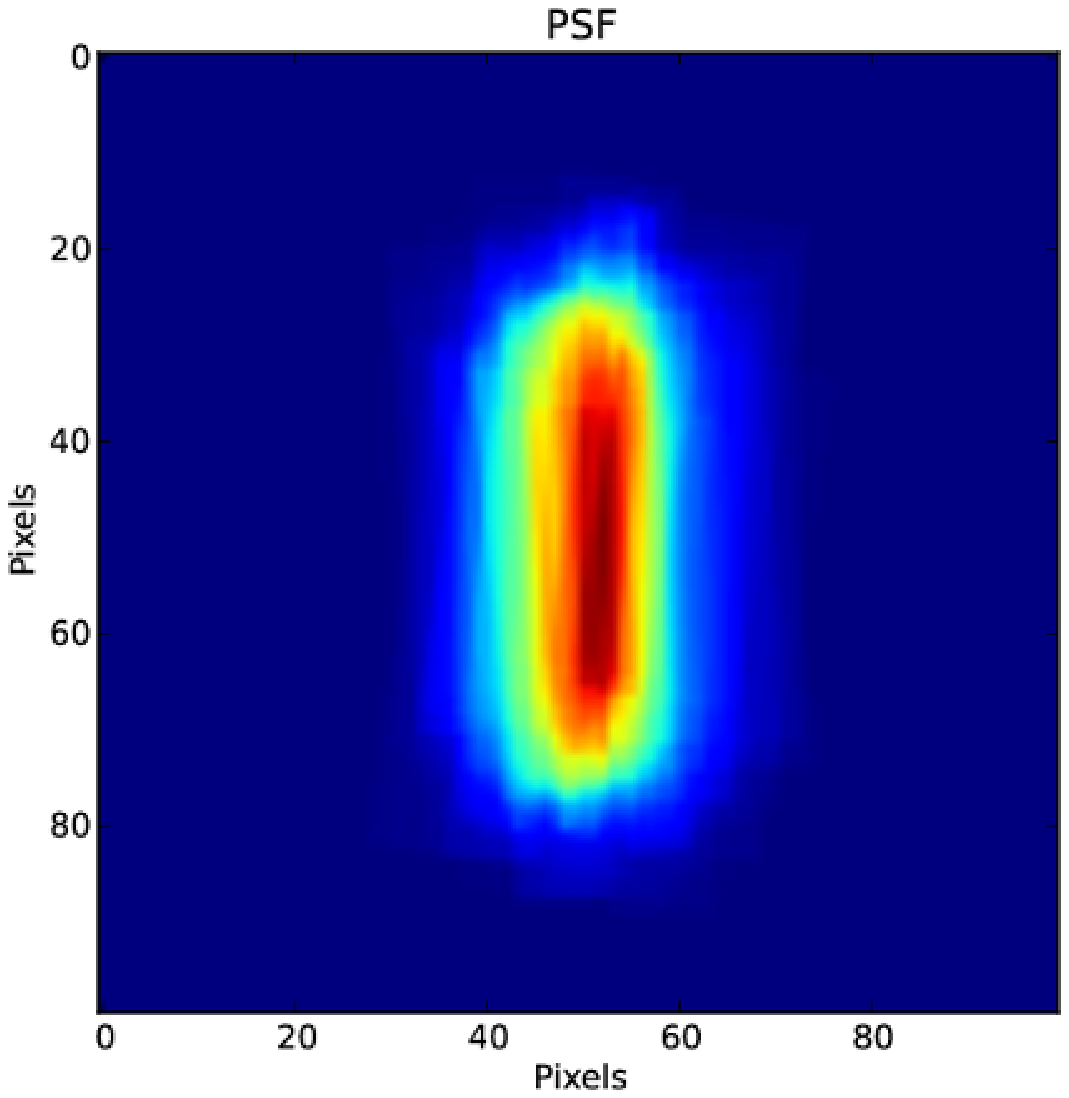}
    
\end{minipage}%
\begin{minipage}{.5\textwidth}
    \centering
    \includegraphics[scale=0.6]{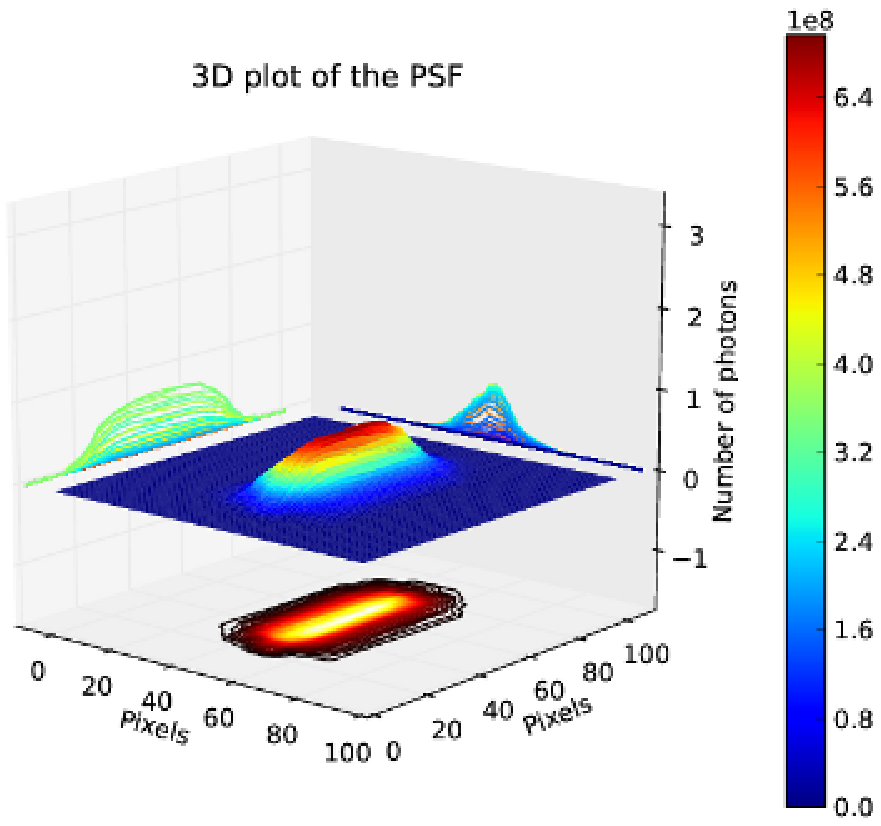} 
    
\end{minipage}
\caption{PSF of the petal made with Ge(111) with a maximum misalignment of 30 arcsec in the 
positioning of the crystal tiles having a maximum radial distortion of 6 meters.}
\label{fig:P_Ge_M30R6}
\end{figure}

We have simulated the petal PSF assuming either bent crystal tiles made of GaAs (220) or bent crystal tiles 
made of Ge(111). The corresponding pass band is either 89--308 keV or 88--290 keV, respectively. 
The resulting PSF is shown in  Figs.~\ref{fig:P_GaAs_M00R0} and \ref{fig:P_GaAs_M30R6} in the case of 
GaAs(220) as the crystal tiles, and in 
Figs.~\ref{fig:P_Ge_M00R0} and \ref{fig:P_Ge_M30R6} in the case of Ge (111).
Figure~\ref{fig:P_GaAs_M00R0} (in case of GaAs(220)) and 
Fig.~\ref{fig:P_Ge_M00R0} 
(in case of Ge(111)) show the PSF when the crystal tiles have no radial distortion and are positioned 
perfectly over the petal frame without any misalignment.
On the other side, Fig.~\ref{fig:P_GaAs_M30R6} and Fig.\ref{fig:P_Ge_M30R6}, respectively for the case of 
GaAs(220) and  Ge(111), show the resulting PSF when the crystal tiles have a radial distortion of 6~m 
(from the required 40~m) and are positioned over the petal frame with a misalignment of 30 arcsec.
The main parameters of the simulated petals are
given in Table~\ref{tab:GaAs_Ge_petal_parameter}.


%

%
%
%
%

\begin{table}[h]
  \begin{center}  
    \begin{tabular}{ l  l  l }
    \toprule
    Parameter 		& \multicolumn{2}{c}{Value} 	\\ \cmidrule(r){2-3}
			& case of GaAs(220) 		& case of Ge(111)	\\ \midrule
    Focal length	& 20 meters 			& 20 meters 		\\
    Energy range	& 89 - 308keV			& 88 - 290keV		\\
    Subtended angle	& 18$\textdegree$ 		& 18$\textdegree$	\\
    No. of Rings	& 33				& 20			\\
    Innermost radius	& 41.71 cm			& 27.67 cm		\\
    Outermost radius	& 137.71 cm 			& 84.67 cm 		\\
    No. of crystal tiles & 913				& 343	 		\\
    Crystal dimension 	& 30 $\times$ 10 $\times$ 2 mm$^3$ 		& 30 $\times$ 10 $\times$ 2 mm$^3$	\\
    Crystal mass (total) & 2.5 g $\times$ 913 = 2282.5 g	& 2.07 g $\times$ 343 = 710 g	\\
    \bottomrule
    \end{tabular}
    \newline
    \caption{Parameters of the simulated petal.}
    \label{tab:GaAs_Ge_petal_parameter}
  \end{center}	
\end{table}


\subsection{Effect of the beam divergence}

\begin{figure}[h!]
  \centering
  \includegraphics[scale=0.3,keepaspectratio=true]{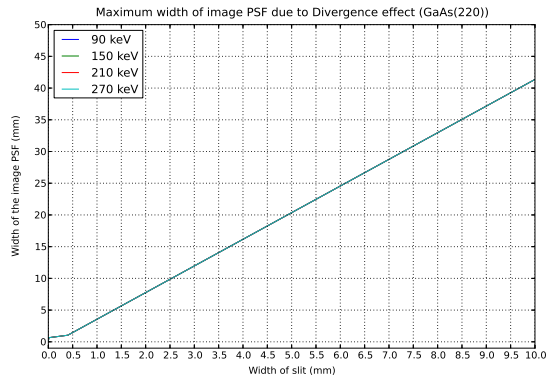} 
  \caption{The effect of the divergence with the varying slit width (on the $x-axis$)
  on the FWHM of the image PSF.}
  \label{fig:divergence}
\end{figure}

A Laue lens for astrophysical observations is built assuming that the incoming 
X-rays are parallel and paraxial. However, in laboratory, it is not possible to achieve this beam configuration given the finite 
distance between source and lens, and to the finite dimension of the source that has a radius of 0.4 mm. In order to 
minimize the beam divergence, a Tungsten bar 20 mm thick with a hole of 3 mm diameter and a lead bar 50 mm thick with a 
hole if 1~mm diameter are placed at the exit window of the X-ray source. The output beam, at a distance of 24.38~m, is further 
collimated with a second lead bar, which has in its center a Tungsten slit 20 mm thick with variable aperture $S_w$. 
After passing through this slit, the beam is incident on the crystal tile, 
which then diffracts and forms an image on the focal plane at a distance 20 meter from the crystal tile. 
The dependence of the Full Width at Half Maximum (FWHM) of the PSF on the slit width $S_w$ is plotted in 
Fig.~\ref{fig:divergence}.

\newpage
\section{Simulating the entire laue lens}

The method adopted to simulate the PSF of the petal is extended to model the entire lens.
The simulation consists of a lens made either with GaAs(220) or with Ge(111) crystal tiles.

\begin{figure}[!ht]
\centering
\begin{minipage}{.5\textwidth}
    \centering
    \includegraphics[scale=0.4]{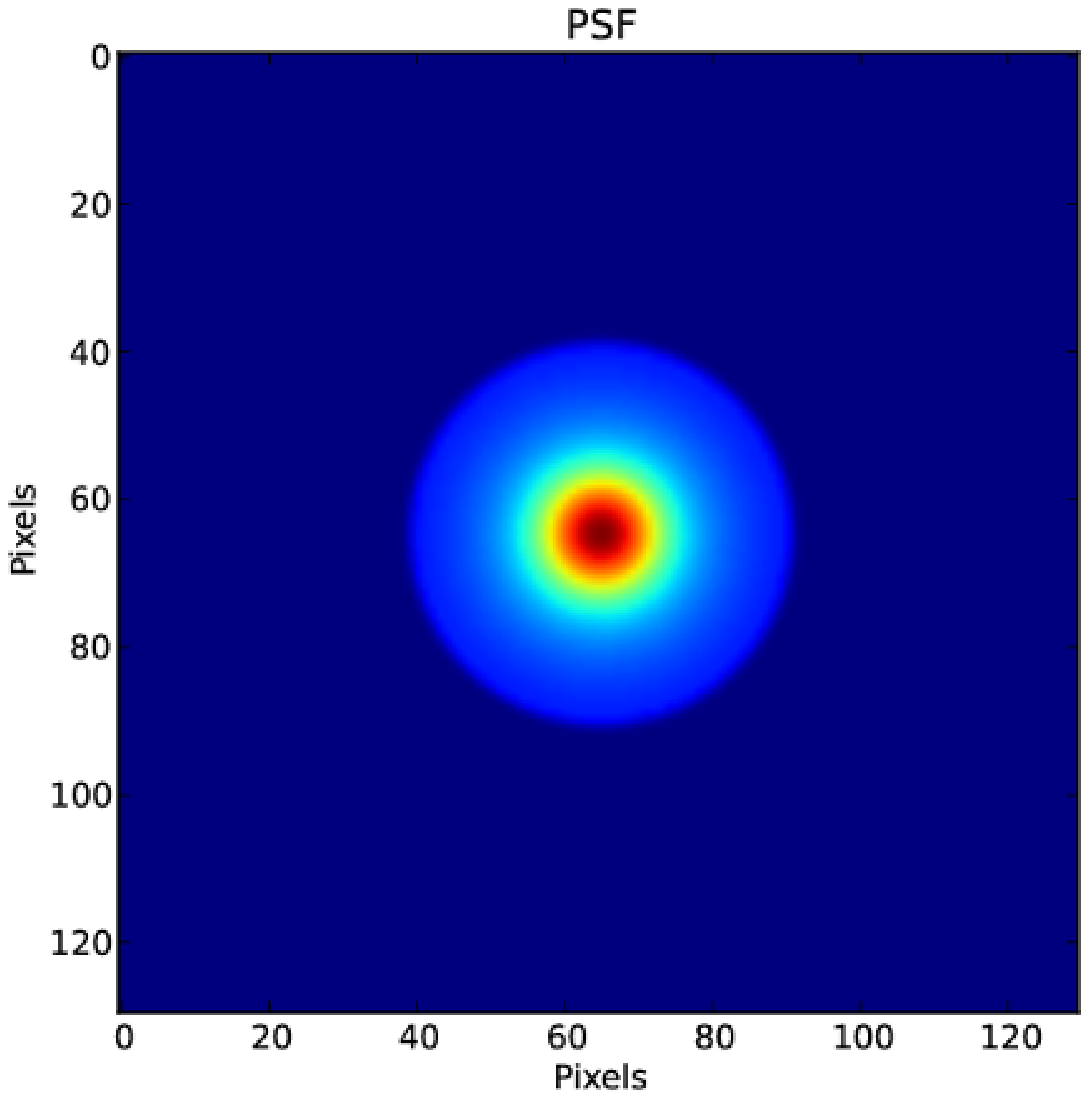}
    
\end{minipage}%
\begin{minipage}{.5\textwidth}
    \centering
    \includegraphics[scale=0.6]{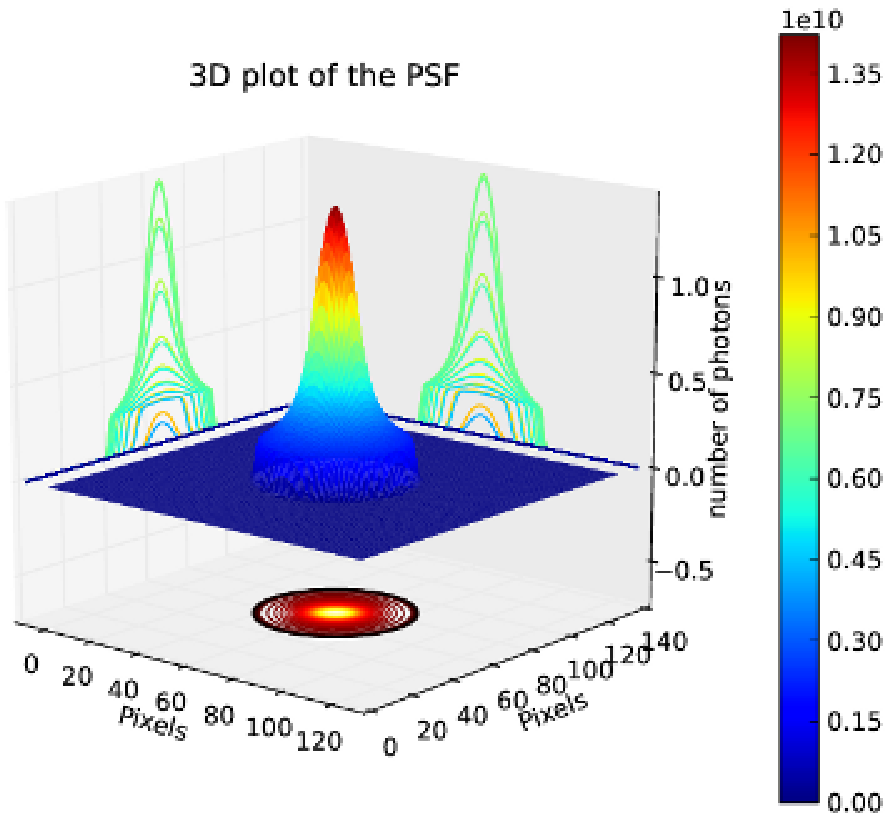} 
    
\end{minipage}
\caption{PSF of the lens made with GaAs(220) without any misalignment errors in the positioning of the 
crystal tiles having no radial distortion.}
\label{fig:L_GaAs_M00R0}
\end{figure}

\begin{figure}[!ht]
\centering
\begin{minipage}{.5\textwidth}
    \centering
    \includegraphics[scale=0.4]{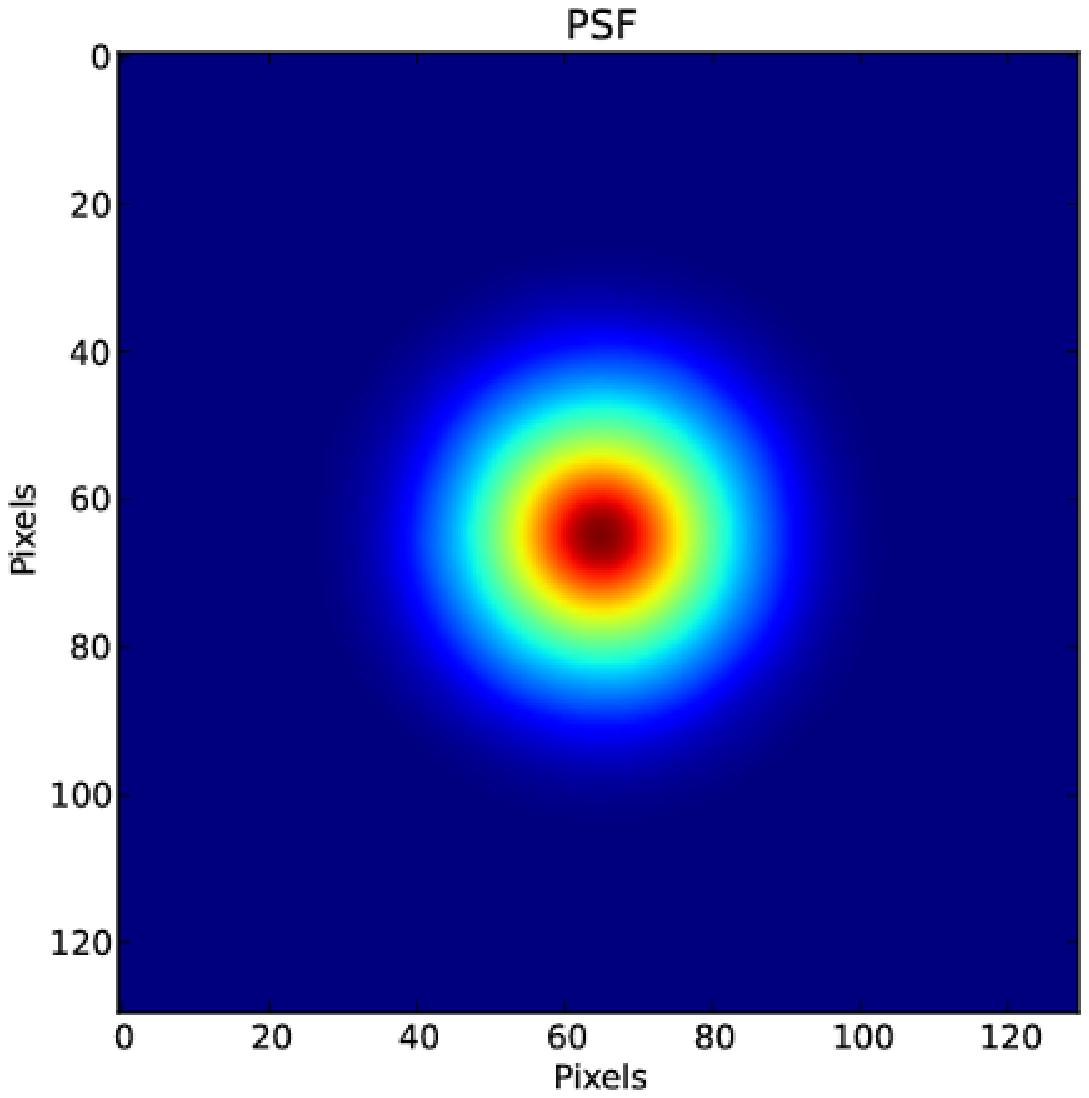}
    
\end{minipage}%
\begin{minipage}{.5\textwidth}
    \centering
    \includegraphics[scale=0.6]{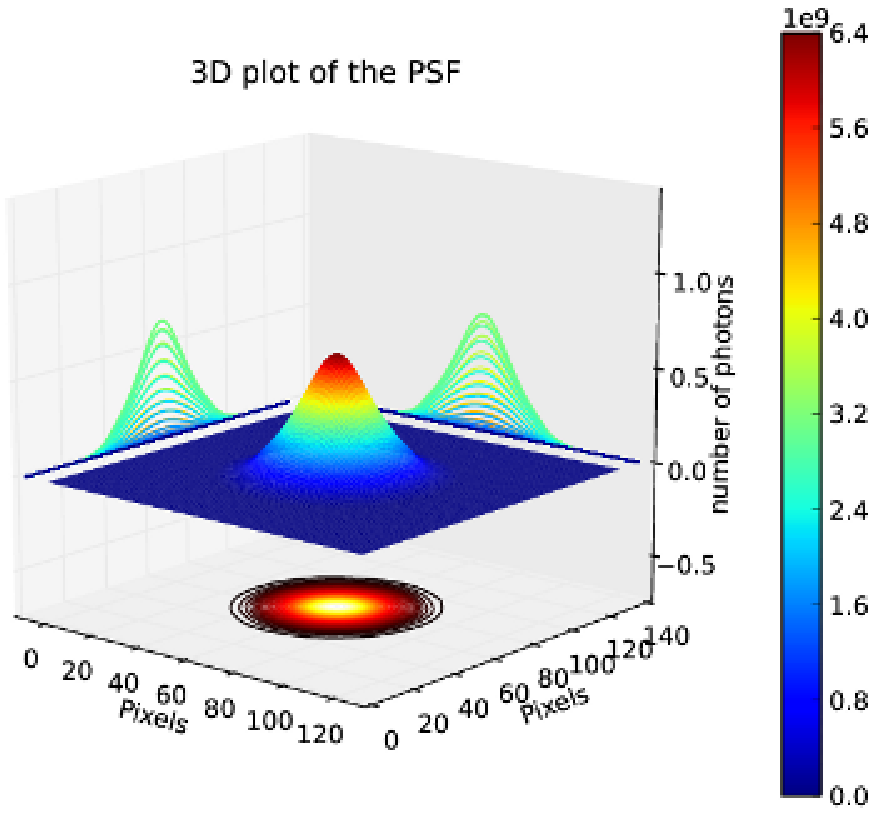} 
    
\end{minipage}
\caption{PSF of the lens made with GaAs(220) with a maximum misalignment of 30 arcsec in the 
positioning of the crystal tiles having a maximum radial distortion of 6 meters.}
\label{fig:L_GaAs_M30R6}
\end{figure}

\begin{figure}[!ht]
\centering
\begin{minipage}{.5\textwidth}
    \centering
    \includegraphics[scale=0.4]{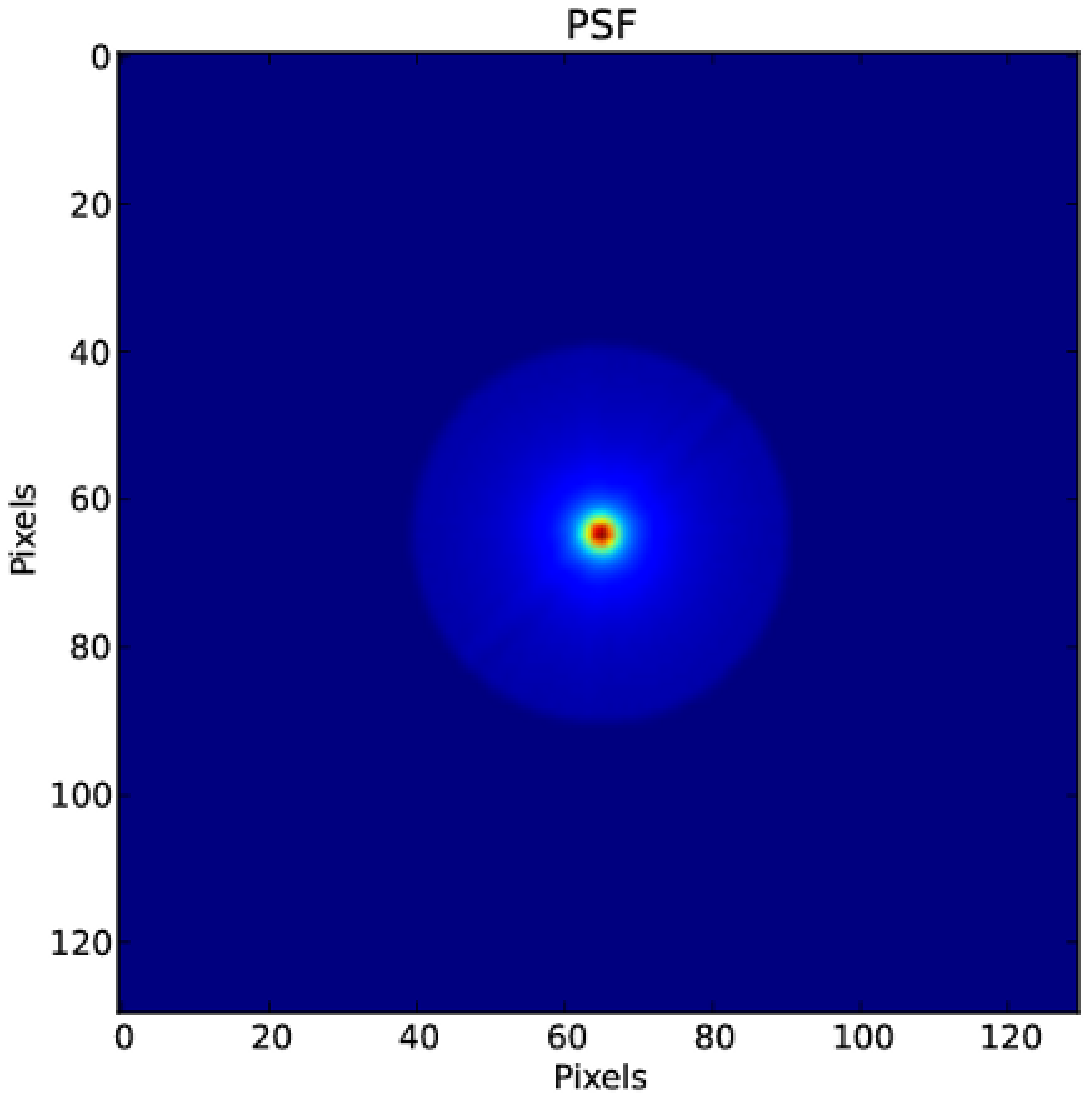}
    
\end{minipage}%
\begin{minipage}{.5\textwidth}
    \centering
    \includegraphics[scale=0.6]{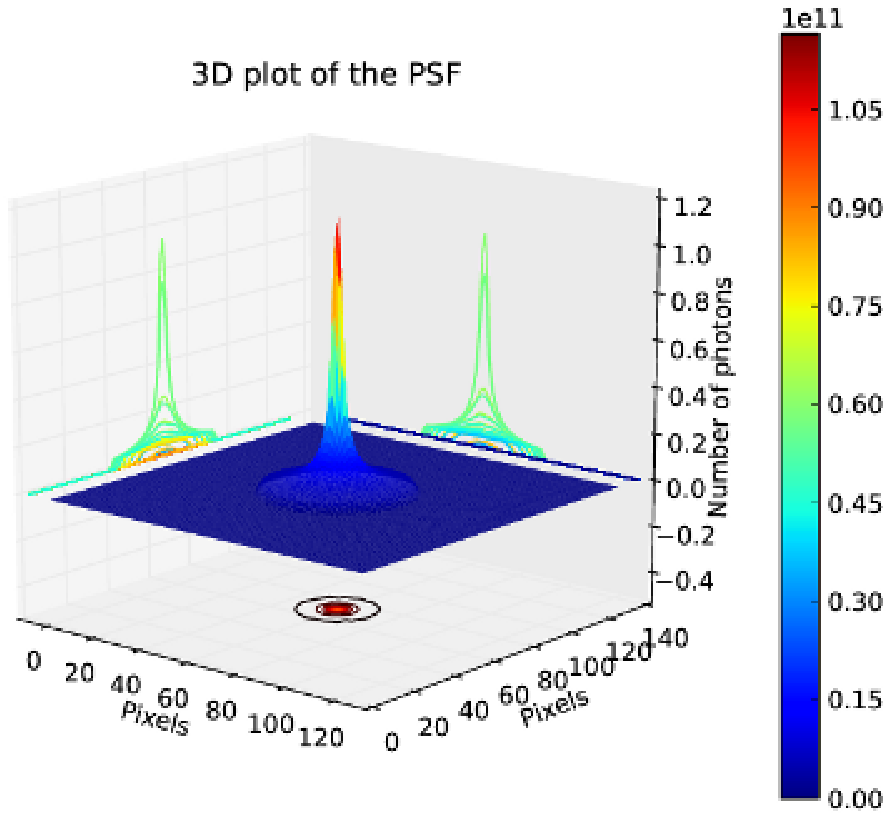} 
    
\end{minipage}
\caption{PSF of the lens made with Ge(111) without any misalignment errors in the positioning of the 
crystal tiles having no radial distortion.}
\label{fig:L_Ge_M00R0}
\end{figure}

\begin{figure}[!ht]
\centering
\begin{minipage}{.5\textwidth}
    \centering
    \includegraphics[scale=0.4]{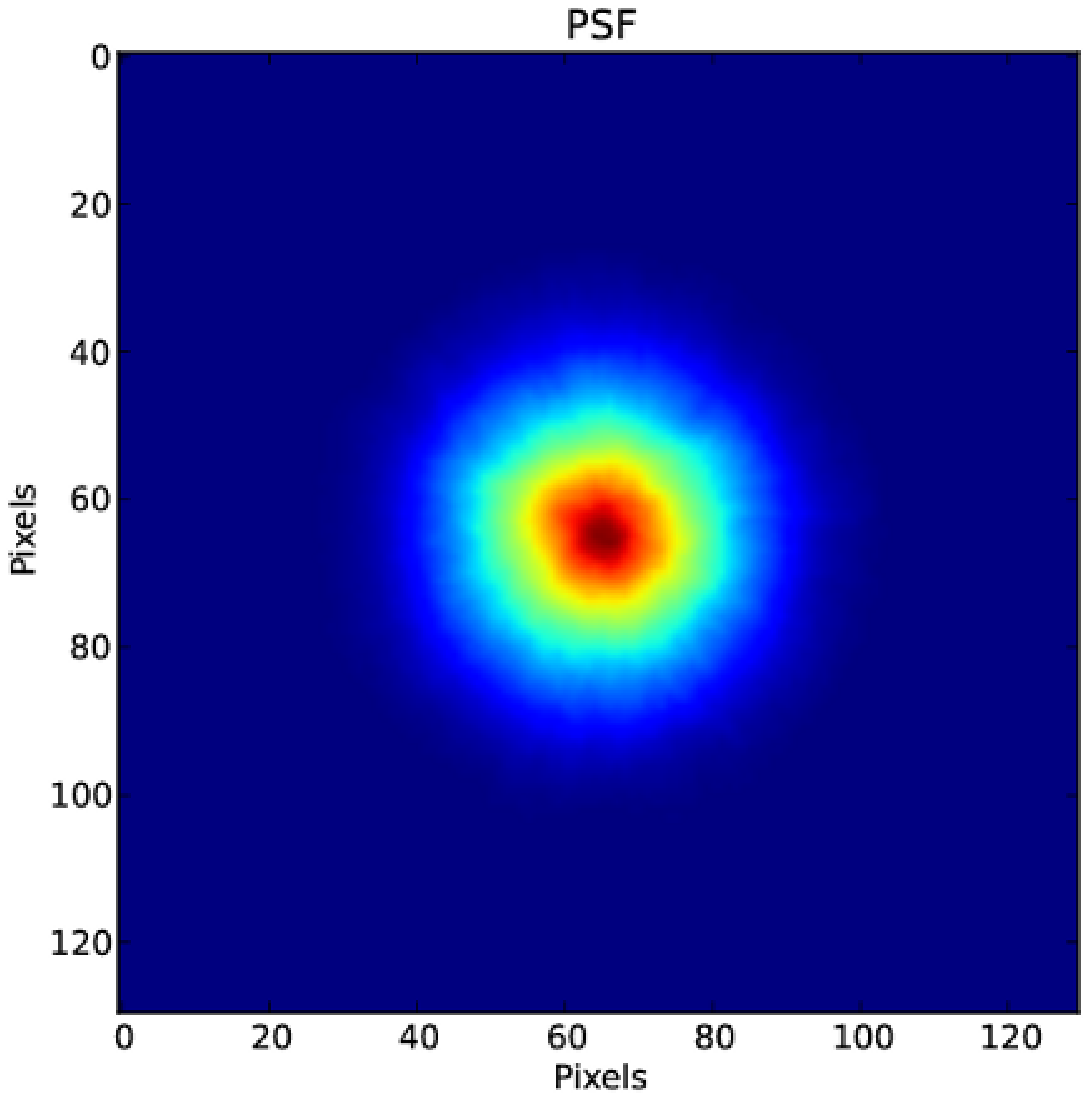}
    
\end{minipage}%
\begin{minipage}{.5\textwidth}
    \centering
    \includegraphics[scale=0.6]{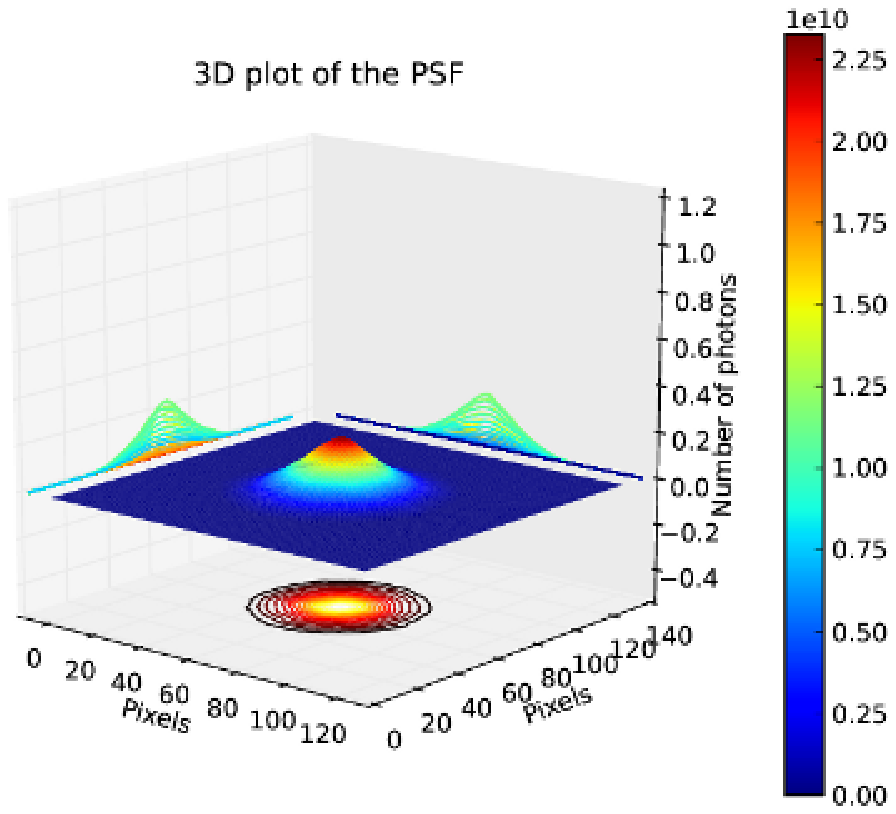} 
    
\end{minipage}
\caption{PSF of the lens made with Ge(111) with a maximum misalignment of 30 arcsec in the 
positioning of the crystal tiles having a maximum radial distortion of 6 meters.}
\label{fig:L_Ge_M30R6}
\end{figure}

The resulting PSF of a lens with pass band from 80 to 600 keV in the case of GaAs (220) is 
shown in Figs.~\ref{fig:L_GaAs_M00R0} and \ref{fig:L_GaAs_M30R6}.
Instead, in the case of  Ge(111) as the crystal tiles, the resulting PSF is shown in Figs.~\ref{fig:L_Ge_M00R0} 
and \ref{fig:L_Ge_M30R6}.

In particular, Fig.~\ref{fig:L_GaAs_M00R0} (in case of GaAs(220) crystal tiles) and Fig.~\ref{fig:L_Ge_M00R0} 
(in case of Ge(111) crystal tiles) show the PSF in the case of no radial distortion and a 
perfect positioning of crystal tiles on the lens frame without any misalignment.
Instead, Fig.~\ref{fig:L_GaAs_M30R6} and Fig.~\ref{fig:L_Ge_M30R6}, respectively, for the 
case of GaAs(220) and  Ge(111), show the PSF when the bent crystal tiles have a 
large radial distortion of 6~m (from the required 40~m) and are positioned on the petal frame 
with a maximum misalignment 
of 30~arcsec.

\begin{table}[h]
  \begin{center}  
  \resizebox{10cm}{!} {
    \begin{tabular}{ l l l }
    \toprule
    Parameter 		& \multicolumn{2}{c}{Value} 	\\ \cmidrule(r){2-3}
			& case of GaAs(220) 		& case of Ge(111)	\\ \midrule	
    Energy range	& 80 - 600keV			& 80 - 600keV		\\
    Focal length	& 20 m			& 20 m		\\
    No. of Rings	& 45				& 28			\\
    Minimum radius	& 20.71 cm			& 12.67 cm		\\
    Maximum radius	& 152.71 cm 			& 93.67 cm		\\
    No. of crystal tiles & 24494			& 9341			\\
    Crystal dimension 	& 30 $\times$ 10 $\times$ 2 mm$^3$		& 30 $\times$ 10 $\times$ 2 mm$^3$		\\
    Crystal mass (total) & 2.5 g $\times$ 24494 = 61.235 kg  	& 2.07 g $\times$ 9341 = 19.335 kg	\\
    \bottomrule
    \end{tabular}
    }
    \newline
    \caption{Parameters of the simulated lens. }
    \label{tab:GaAs_Ge_lens_parameter}
  \end{center}	
\end{table}

The values of the main parameters of the lens are
given in Table~\ref{tab:GaAs_Ge_lens_parameter}.


\newpage

\section{Discussion}

In this section we discuss the lens simulation results in the case of GaAs(220) and Ge(111) crystal tiles, 
and their consequences.

\subsection{FWHM profile}
\begin{figure}[h!]
  \centering
  \includegraphics[scale=0.3,keepaspectratio=true]{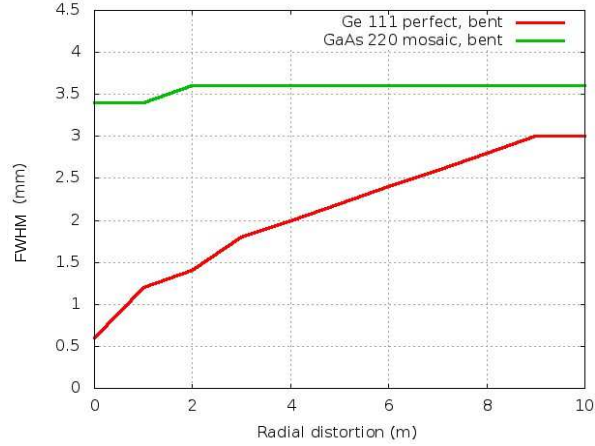} 
  \caption{FWHM profile of the lens made with perfect positioning of the crystal tiles
  on the lens frame.}
  \label{fig:L_GaAs_Ge_FWHM}
\end{figure}

Any misalignment in the positioning of the crystal on the lens/petal frame,
and also any distortion in the curvature radius will affect the FWHM of the PSF.
Fig.~\ref{fig:L_GaAs_Ge_FWHM} shows the dependence of FWHM on the radial distortion for the two lens cases, 
assuming a perfect positioning of the crystal tiles on the lens frame. 
As it can be seen, when there is no misalignment and no radial distortion, the FWHM is 3.4 mm in the case of 
GaAs(220) and 0.6 mm for Ge(111) as the crystal tile used for building the lens.

\subsection{Peak intensity profile}

\begin{figure}[!ht]
\centering
\begin{minipage}{.5\textwidth}
    \centering
    \includegraphics[scale=0.3,keepaspectratio=true]{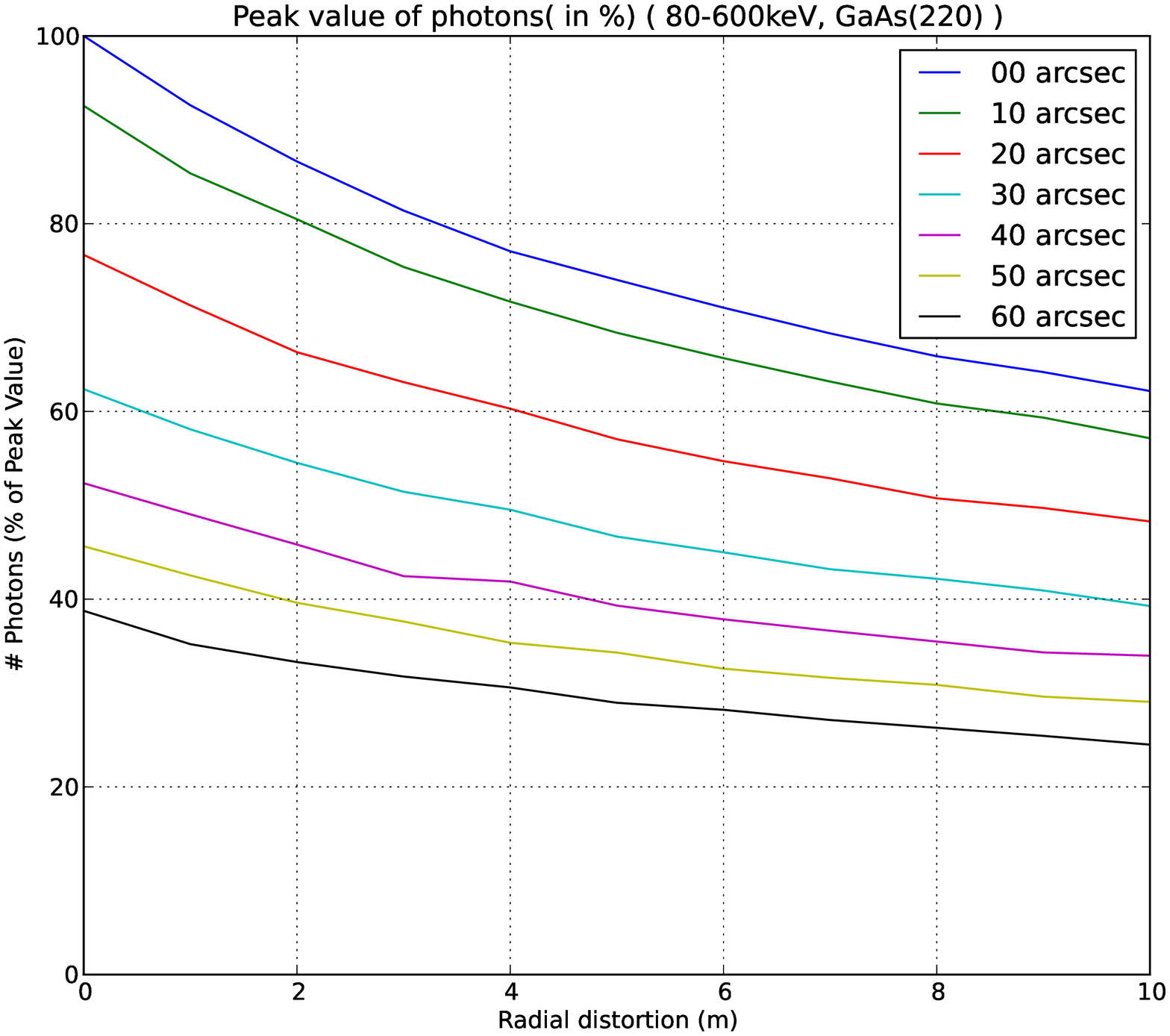} 
    
\end{minipage}%
\begin{minipage}{.5\textwidth}
    \centering
    \includegraphics[scale=0.4,keepaspectratio=true]{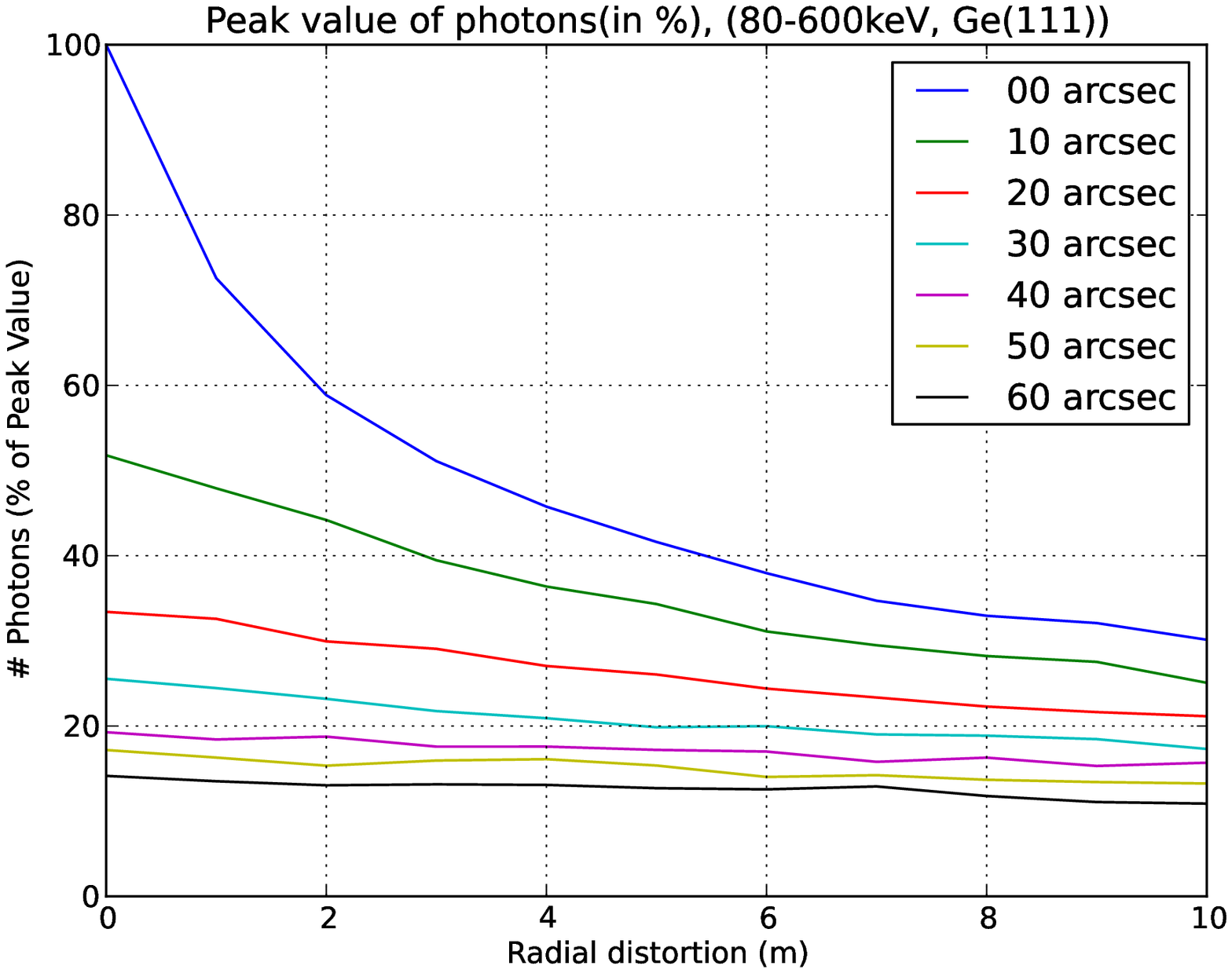} 
    
\end{minipage}
\caption{Peak intensity profile of the lens made with GaAs(220) (\textit{left}) and Ge(111) (\textit{right})
for different values of crystal misalignment and radial distortion. 
The values in the $legend$ shows maximum misalignment (in arcsec) in the positioning of crystals.}
\label{fig:L_GaAs_Ge_PI}
\end{figure}

The effect of misalignment on the peak intensity is plotted in Fig.~\ref{fig:L_GaAs_Ge_PI}.
The peak intensity, as expected, gets reduced with the increase in misalignment error and
radial distortion. Assuming 100\%, the peak intensity of a lens made by perfect positioning of GaAs(220) 
crystal tiles 
and no radial distortion, this value  decreases to about 55\% for a misalignment
(in the positioning of the crystal tile on the lens frame) of 30 arcsec and a large radial distortion of 
6 meters.
In the case of Ge(111), the corresponding reduction is
of about 80\%.

\subsection{Radial profile}
\begin{figure}[h!]
\centering
\begin{minipage}{.5\textwidth}
  \includegraphics[scale=0.25,keepaspectratio=true]{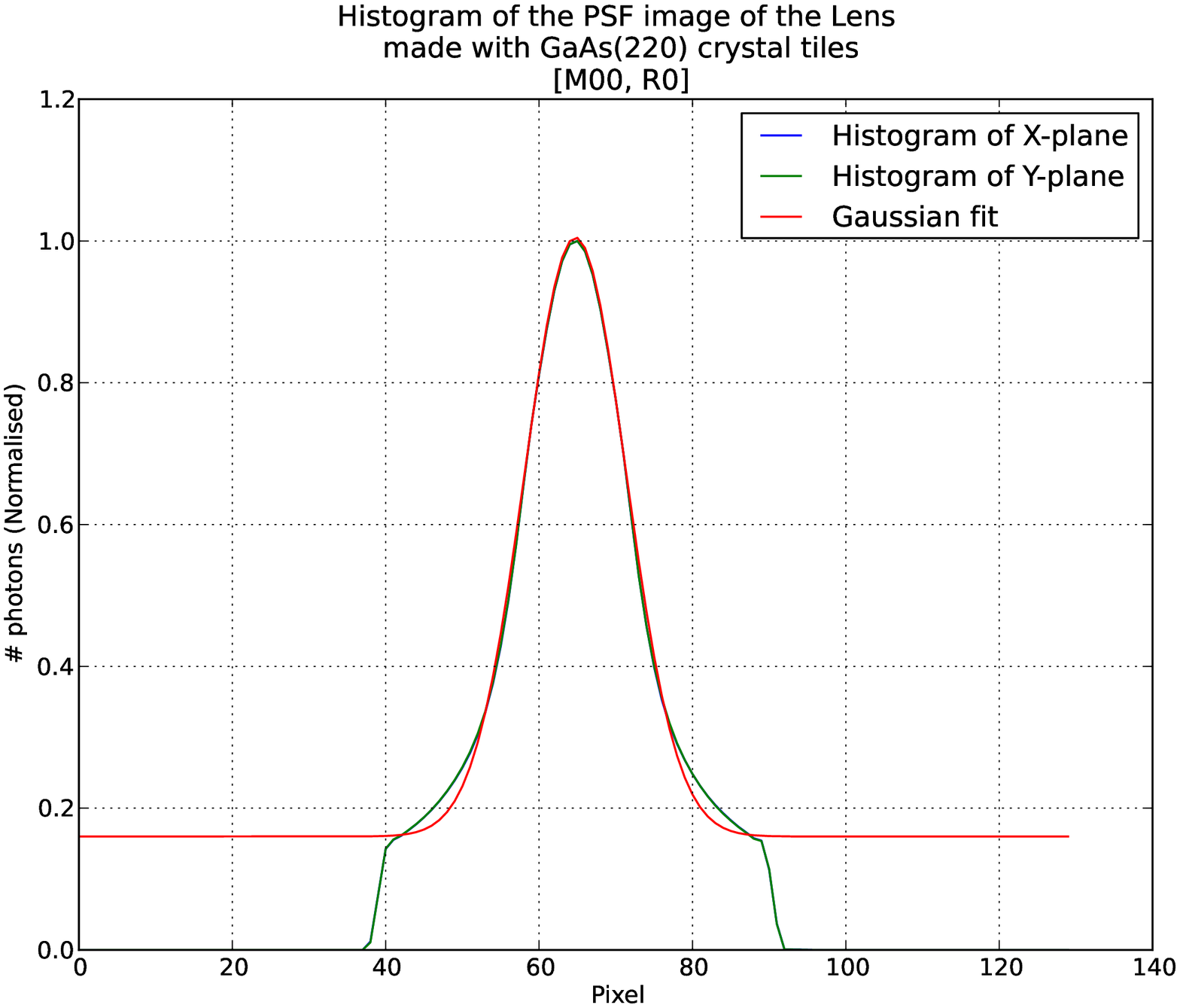}
\end{minipage}%
\begin{minipage}{.5\textwidth}
  \includegraphics[scale=0.25,keepaspectratio=true]{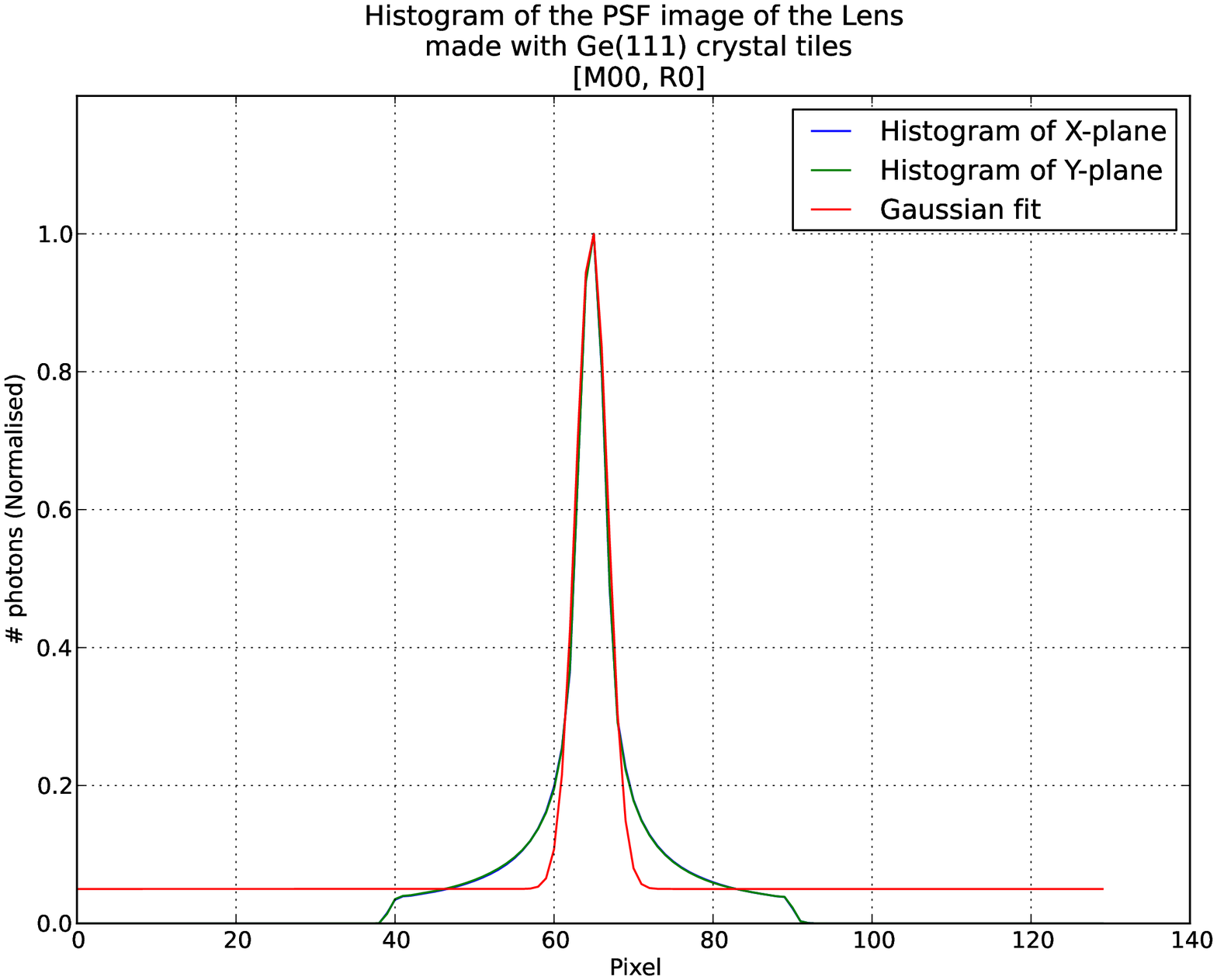} 
    
\end{minipage}
  \caption{Histogram of the image PSF of the lens made by GaAs(220) crystal tiles (\textit{left}), 
  and that by Ge(111) crystal tiles (\textit{right}). 
  This lens is made with perfect positioning of the crystal tile, 
  having no radial distortions.}
  \label{fig:histo_Lens}
\end{figure}

The radial profile of the PSF is the cumulative distribution of the number of photons collected 
along the radial distance from the focal point.
The PSF of the lens with a perfect alignment 
(misalignment = 0.0 arcsec $[M00]$) of the crystal tiles and no radial distortion of the bent crystals
(radial distortion = 0 meters $[R0]$) is shown in Fig.~\ref{fig:histo_Lens} 
(case of the lens made by GaAs(220) crystal tiles on the \textit{left} 
and that of Ge(111) crystal tiles on the \textit{right}).

\begin{figure}[h!]
  \centering
  \includegraphics[scale=0.7,keepaspectratio=true]{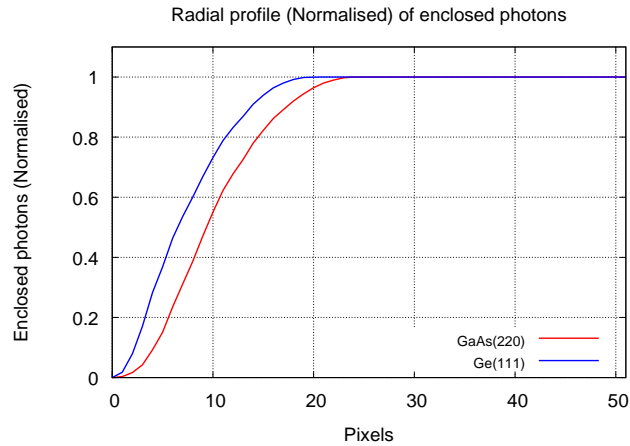}
  \caption{Normalised radial profile of the lens made with perfect positioning of the crystal tile, 
  having no radial distortions. This profile has been calculated by neglecting the 
  'trapezoidal' base of the PSF image.}
  \label{fig:Normalised_radial_profile_Lens}
\end{figure}

As it can be seen from the figure, the PSF does not show a  `Gaussian` shape, but has at the bottom 
a 'trapezoidal'
shape. This shape introduces an offset and increases the background level of the 'signal'. 
The normalised radial profiles 
obtained by subtracting this offset is plotted in 
Fig.~\ref{fig:Normalised_radial_profile_Lens}.
The pixel dimension is 200 $\mu$m $\times$ 200 $\mu$m.

\subsection{Effective area}

The effective area at an energy $E$ is the product of the geometrical area of the optics times
its reflection efficiency at the same energy $E$.
When the entire energy range (90-600 keV) of the lens is divided into 10 equal bins (in logarithmic scale),
the total geometric area in each bin $GA_{total}^{bin}$ is taken as the surface area of the lens 
corresponding to the energy range of each binning.

\begin{equation}
  GA_{total}^{bin} = N_c(\Delta E) \times xtal_{area}
\end{equation}

where $N_c(\Delta E)$ is the number of crystal tiles associated with the corresponding energy range of the bin,
and $xtal_{area}$ is the surface area of a single crystal tile.
The effective area $Area_{eff}^{bin}$, in each binning is given by:

\begin{equation}
  Area_{eff}^{bin} = GA_{total}^{bin} \times \overline{R_{bin}}
  \label{eqn:EA_method2}
\end{equation}

where R$_{bin}$ is the mean reflectivity in each bin.

\begin{figure}[!ht]
\centering
\begin{minipage}{.5\textwidth}
    \centering
    \includegraphics[scale=0.65,keepaspectratio=true]{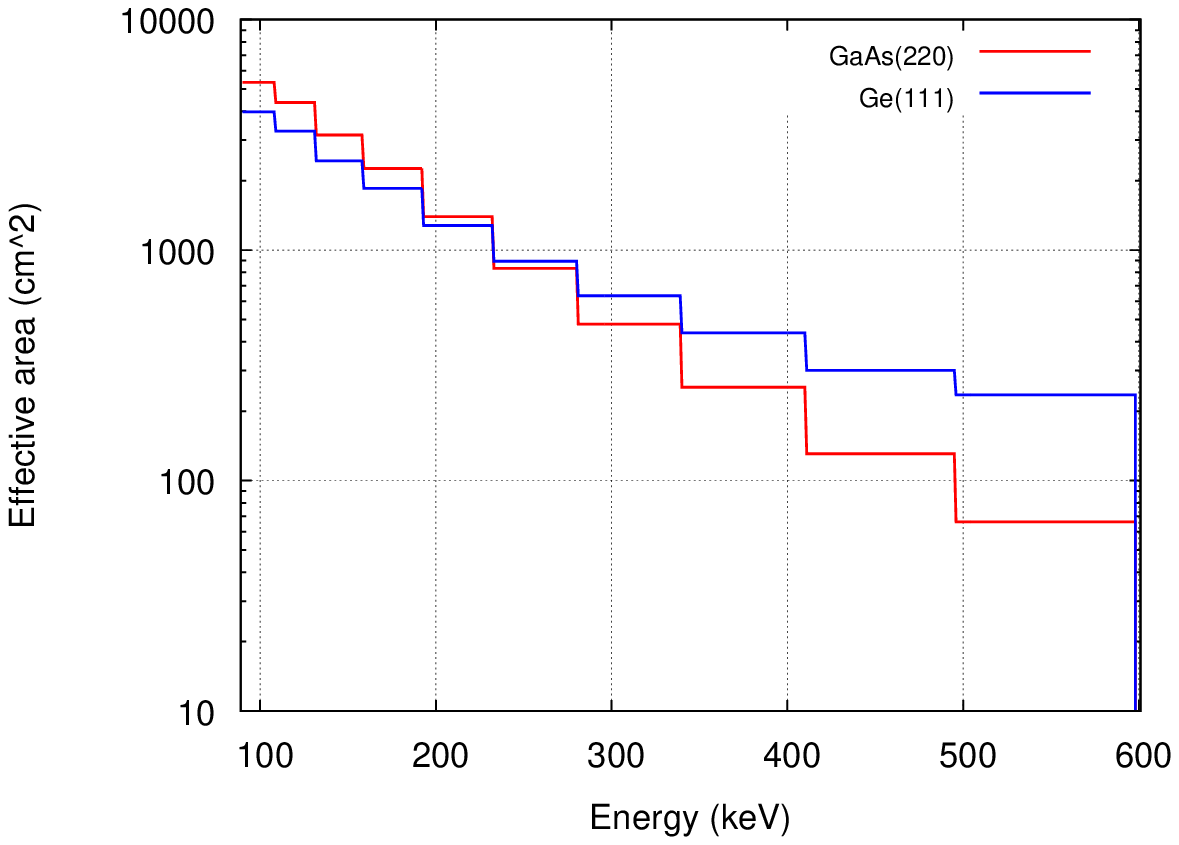}
    
\end{minipage}%
\begin{minipage}{.5\textwidth}
    \centering
    \includegraphics[scale=0.75,keepaspectratio=true]{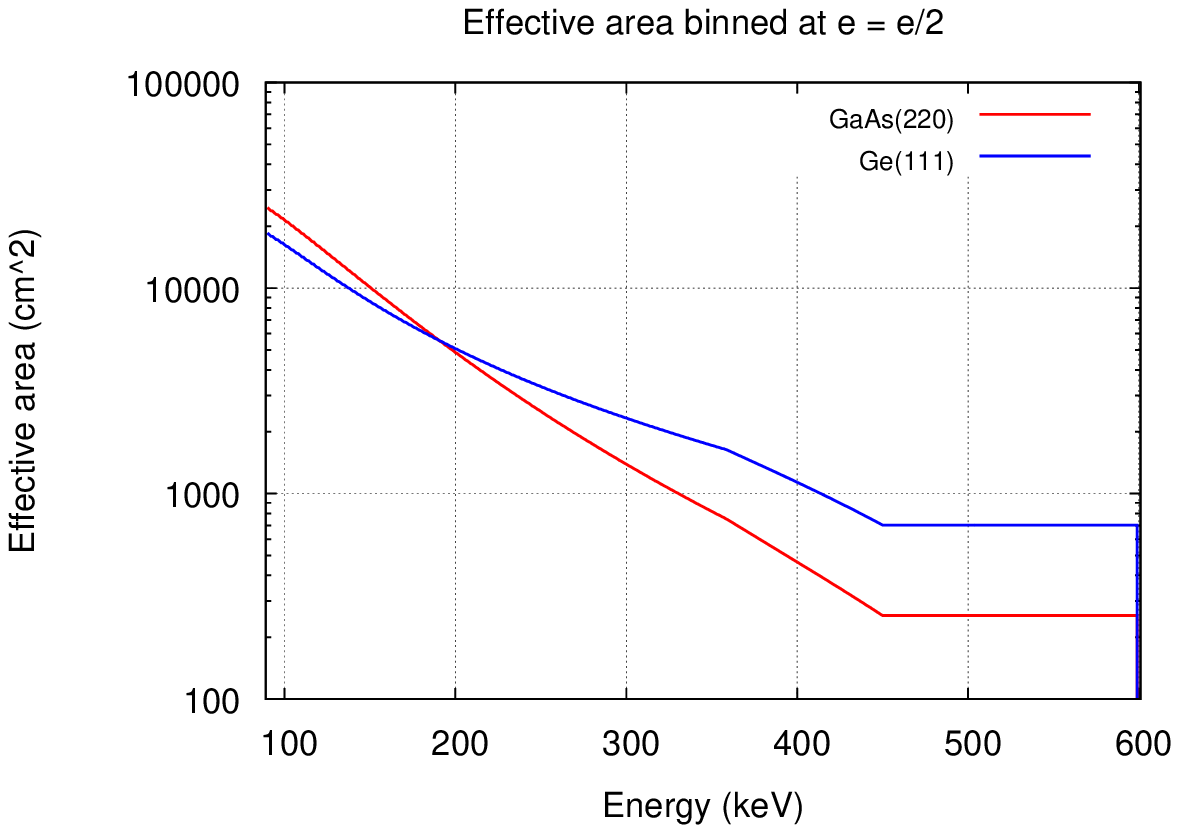} 
    
\end{minipage}
\caption{Effective area of the Lens made of GaAs(220) as well as Ge(111) crystal tiles
  calculated for 10 bins sampled equally in logarithmic scale (\textit{left}) 
  and for $\Delta E = e/2 $ \textit{right}.}
\label{fig:EA_Gaas_Ge}
\end{figure}

Fig.~\ref{fig:EA_Gaas_Ge} shows the effective area calculated from Eq.~\ref{eqn:EA_method2} 
in case of GaAs(220) and Ge(111), respectively.
The large values of effective area at lower energies and smaller values at higher energies
is mainly due to the large difference in the number of crystal tiles corresponding to those energies.
For example, the number of Ge(111) crystal tiles corresponding to lowest energy bin are around 2092, 
but there are only 68 Ge(111) crystal tiles corresponding to the highest energy bin. 
For lens made with GaAs(220) crystal tiles there are 5580 tiles for the lowest energy bin
and 182 tiles for the highest energy bin.

\subsection{Continuum sensitivity}

Sensitivity of a telescope is defined as the minimum intensity, $I^{min}_s$ which can be 
"detected" by the detector in an observation time $T_{obs}$.
For a focusing telescope, if the total noise measured is only due to the background 
(the noise due the source is negligible), 
the sensitivity, $I^{min}_{ft}(E)$ is given by:

\begin{equation}
  I^{min}_{ft}(E) = n_{\sigma}\frac{\sqrt{B(E)} \sqrt{A_d}}{\eta_{d} f_{\epsilon} A_{eff} \sqrt{\Delta E} \sqrt{T_{obs}}} 
 \label{eqn:CS_ft}
\end{equation}

where, 

\begin{itemize}
  \item $n_\sigma$ is the confidence level (usually 3$\sigma$);
  \item $B(E)$ is the intensity of the measured background spectrum (in $counts/s/cm^2/keV$) at the energy $E$;
  \item $\eta_{d}$ is the efficiency of the position sensitive detector;
  \item $A_d ( = \pi R_{spot}^2)$ is the detector area covered by the incoming photons within a radius of $R_{spot}$;
  \item $\Delta E$ is the energy band around $E$;
  \item $T_{obs}$ is the observation time;
  \item $f_{\epsilon}$ is the fraction of photons that is incident on the detector area $A_d$;
  \item $A_{eff}$ is the effective area at an energy $E$ of the telescope optics, given by the product of the
  geometrical area of the optics times its reflection efficiency at that energy $E$.
\end{itemize}

\begin{figure}[h!]
  \centering
  \includegraphics[scale=0.9,keepaspectratio=true]{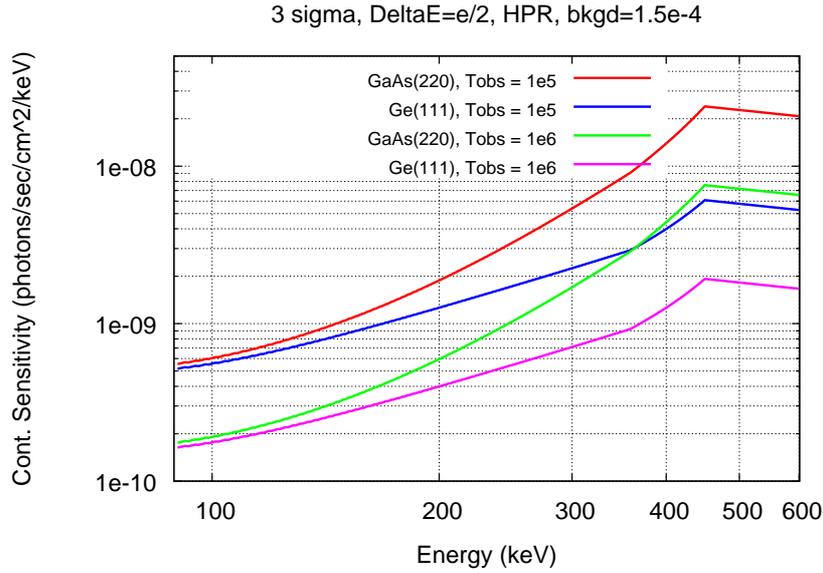}
  \caption{$3 \sigma$ Continuum sensitivity of the Lens with time of observation $10^5$ seconds 
  and $10^6$ seconds with $\Delta E = e/2$, background = $1.5 \times 10^{-4}$ and $R_{spot} =$ HPR }
  \label{fig:CS_var_Tobs_eby2}
\end{figure}

Figure~\ref{fig:CS_var_Tobs_eby2} shows the 3 $\sigma$  
sensitivity for an observation time of 10$^5$ s and 10$^6$ s, along with $\Delta$E = E/2 and 
background = 1.5 $\times$ 10$^{-4}$ counts/sec/cm$^2$/keV.

\subsection{Line sensitivity}

The line sensitivity can be obtained from  the continuum sensitivity by superposing an 
emission line to the continuum intensity.
In the case of a focusing telescope, if the source continuum level can be accurately determined,
the minimum detectable intensity I$_L^{min}$ (in photons/s/cm$^2$),
of a line is given by

\begin{equation}
  I_L^{min}(E_L) = 1.31 n_{\sigma}\frac{\sqrt{[2 B(E_L) A_d + I_c(E_L) \eta_{d} f_{\epsilon} A_{eff}]\Delta E} } { \eta_{d} f_{\epsilon} A_{eff} \sqrt{T_{obs}}} 
 \label{eqn:LS_ft}
\end{equation}

where, 

\begin{itemize}
  \item $n_\sigma$ is the confidence level (usually 3$\sigma$);
  \item $B(E_L)$ is the intensity of the measured background spectrum (in counts/s/cm$^2$/keV) at the energy $E_L$;
  \item $A_d ( = \pi R_{spot}^2)$ is the detector area covered by the incoming photons within a radius of $R_{spot}$;
  \item $I_c(E_L)$ is the intensity (in counts/s/cm$^2$/keV) at the continuum of the source at the centroid of the line;
  \item $\eta_{d}$ is the efficiency of the position sensitive detector;
  \item $f_{\epsilon}$ is the fraction of photons that is incident on the detector area A$_d$;
  \item $\Delta E$ is the FWHM around $E_L$. This value depends upon the energy resolution of the detector;
  \item $T_{obs}$ is the observation time;
  \item $A_{eff}$ is the effective area at an energy $E_L$ of the telescope optics, given by the product of the 
  geometrical area of the optics times its reflection efficiency at that energy E$_L$.
\end{itemize}

\begin{figure}[h!]
  \centering
  \includegraphics[scale=0.85,keepaspectratio=true]{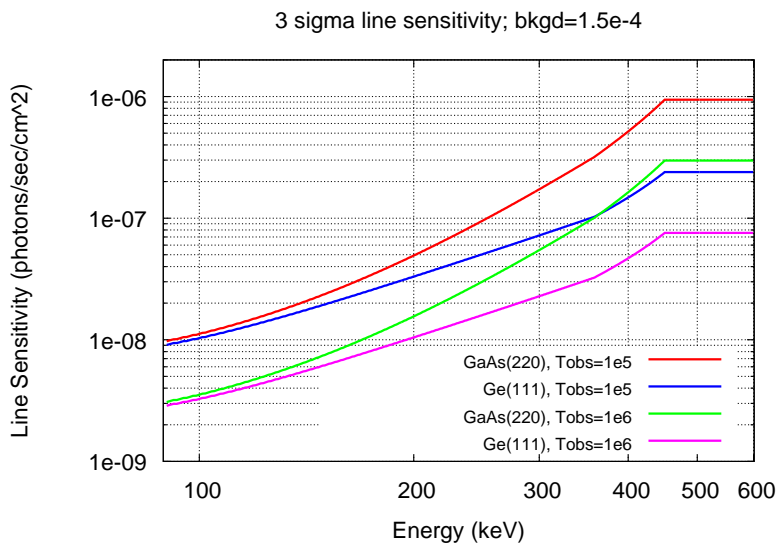}
  \caption{Line sensitivity of the lens.}
  \label{fig:LS_Lens_eby2}
\end{figure}

The line sensitivity for the lens made with perfectly positioned GaAs(220) or Ge(111) bent
crystal tiles with no radial distortion, is shown in 
Fig.~\ref{fig:LS_Lens_eby2}. The sensitivity of each lens is plotted 
separately for 10$^5$ s and 10$^6$ s observation time. The values of other parameters are the following:
n$_\sigma$ = 3, $A_d$ is the area corresponding to half power radius 
(for $f_{\epsilon}$ = 50\% of enclosed photons) 
$\eta_{d}$ = 90\%, $\Delta E$ = 2 keV, which is the 
energy resolution of the CZT detector intended for our use, and finally A$_{eff}$ corresponds to 
the data from Fig.~\ref{fig:EA_Gaas_Ge} with a filling factor of 0.2 mm.

\section{Conclusion}
We have presented the performance of a simulated lens petal and of a lens 
made of either GaAs(220) or Ge(111) crystal tiles. 
The effect of divergence of the beam on the petal PSF has been simulated.
The dimension of the diffracted image of GaAs(220) crystal tile glued on the 
petal structure is found to be similar to that calculated through simulation, including the divergence.
The divergence of the beam is a significant factor affecting the PSF.

\begin{table}[h]
  \begin{center}  
    \begin{tabular}{ c c c c c }
    \toprule
    Energy & \multicolumn{2}{c}{Simulation ($T_{obs}=10^5$s)} & ISGRI 	& SPI \\ \cmidrule(r){2-3}
    (keV)  & GaAs(220) 		    & Ge(111) 		     & $T_{obs}$=10$^5$s   &$T_{obs}$=10$^6$s \\ \midrule	
    100	   & 6.04 $\times$ 10$^{-10}$ & 5.57 $\times$ 10$^{-10}$ & 2.85 $\times$ 10$^{-6}$ & 7.0 $\times$ 10$^{-6}$ \\
    500	   & 2.24 $\times$ 10$^{-8}$  & 5.76 $\times$ 10$^{-9}$  & 9.83 $\times$ 10$^{-6}$ & 1.5 $\times$ 10$^{-6}$ \\
    \bottomrule
    \end{tabular}
    \newline
    \caption{Comparison of the 3$\sigma$ continuum sensitivity (in photons/sec/cm$^2$/keV) of simulated lens 
    with ISGRI and SPI on-board INTEGRAL.}
    \label{tab:CS_comparison_INTEGRAL}
  \end{center}	
\end{table}

The simulation of the petal has been extended to model an entire lens made of petals.
The effective area, the continuum sensitivity as well as the line sensitivity are calculated.
These results show that this lens is about 3 orders of magnitude more sensitive 
(see Table \ref{tab:CS_comparison_INTEGRAL}) than ISGRI\cite{INTEGRAL_IBIS} 
and SPI\cite{INTEGRAL_SPI} on-board INTEGRAL satellite.

\acknowledgments     
 
The authors wish acknowledge the financial 
support by the Italian Space Agency (ASI) through the project ``LAUE: Una lente per i raggi Gamma''
Under contract I/068/09/0.
Vineeth Valsan and Vincenzo Liccardo are supported by the
Erasmus Mundus Joint Doctorate Program by Grant Number 2010-1816 from the EACEA of the European
Commission.

\bibliography{RE_1}
\bibliographystyle{spiebib}

\end{document}